# Weakly nonlinear analysis for car-following model with consideration of cooperation and time delays

Dihua Sun [a,b] Dong Chen[a,b*] Min Zhao [a,b] Yuchu He [a,b] Hui Liu [c]

[a] Key Laboratory of Dependable Service Computing in Cyber Physical Society of Ministry of Education, Chongqing University, Chongqing 400044, China

[b] School of Automation, Chongqing University, Chongqing 400044, China

[c] College of Mechanical and Electrical Engineering, Chongqing University of Arts and Sciences, Chongqing 402160, China

**Abstract**

In traffic systems, cooperative driving has attracted the researchers' attentions. A lot of works attempt to understand the effects of cooperative driving behavior and/or time delays on traffic flow dynamics for specific traffic flow model. This paper is a new attempt to investigate analyses of linear stability and weak nonlinear for the general car-following model with consideration of cooperation and time delays in cyber-physical perspective. We derive the linear stability condition and study that how the combinations of cooperation and time delays affect the stability of traffic flow. Burgers equation and Korteweg de Vries (KdV) equation for the generic car-following model considering cooperation and time delays are derived. Their solitary wave solutions and constraint conditions are concluded. We investigate the property of cooperative optimal velocity(OV) model which estimates the combinations of cooperation and time delays about the evolution of traffic waves using both analytic and numerical methods. The results indicate that delays and cooperation are model-dependent, and cooperative behavior could inhibit the stabilization of traffic flow. Moreover, delays of sensing to relative motion are easy to trigger the traffic waves; delays of sensing to host vehicle are beneficial to relieve the instability effect a certain extent.

## 1. Introduction

The issues of cooperative traffic systems have been widely investigated in recent years. The vehicles can exchange information with other vehicles and road-side units through vehicle-to-X (V2X) communication including vehicle-to-vehicle (V2V) and vehicle-to-infrastructure (V2I), and hence may exhibit luxuriant cooperative dynamic behavior compared to traditional vehicles that do not communicate [1-2]. For purpose of promoting traffic safety, efficiency, green and comfort, transportation cyber-physical systems (T-CPS) have been proposed by some researchers [3-6]. In this paper, we focus on cooperative traffic systems using V2V communication.

From different views of research, there are two main aspects of cooperative traffic systems in the state-of-the-art methods: traffic flow theory and traffic control. One main line of traffic systems has direct connections to develop traffic control strategies such as constructing decentralized control [7], cooperative adaptive cruise control(CACC) [1, 8-9], sliding-mode control [10] and gain scheduling techniques[11]. However, we care about the complex cooperative dynamics of traffic in this paper.

To solve complex traffic problems, various traffic flow models such as car-following models [12-19], lattice hydrodynamic models [20-26] and macroscopic traffic flow model [27-31] have been investigated. Some complex mechanisms behind the phenomena of traffic flow have been revealed in the microscopic and macroscopic levels. A lot of effort attempted to understand the characteristics of traffic flow dynamics for specific traffic flow model.

Stability is one of key factor for the characteristics of traffic flow, which focuses on steady state under small perturbations [12-15, 18]. Instabilities of traffic flow resulting in traffic waves are caused by delays in stimulating acceleration or adapting the speed to the actual traffic conditions [32-33]. These delays do not only the consequence of finite acceleration and braking capabilities, but also result from finite reaction times of the drivers [34-35]. If traffic density is sufficiently high and more than the critical value, delays lead to a positive feedback on density and speed perturbations and lead to phantom jams [36]. It is important to reveal that there exist various complex instability mechanisms in traffic flow due to delays. Delays in time-continuous models have been studied dated from 1961year. Car-following models have found widespread application and become rather more complicated than their classical predecessors. Bando et al. [32] proposed OV model with time delays and rediscovered a rich source of dynamical behavior. Car-following model considering driver's reaction time was put forward to reveal the oscillations in vehicle velocity induced by encountering slower vehicles [33]. The local and global bifurcations of car-following model with delays are investigated [34] and different periodic bifurcations of traffic state were analyzed. Delays, inaccuracies and anticipation in microscopic traffic models were systematically investigated [35]. Two causes for the instability of the traffic flow were studied [37-38]: the time lag caused by finite engine performance and the delays caused by the finite reaction times of the drivers. Some researchers focused on delays in sensing headway and velocity and then analyzed the impact of traffic jams [39]. Those contributions investigated that delays affects the evolution of traffic flow from different perspectives.

* Corresponding author

Cooperation can refer to the driving behavior that includes driver's responses to multiple vehicles ahead/behind in traffic flow via V2V. It is well-known that cooperation increases the stability domain of traffic flow [40-42] and helps stabilize the evolution of the disturbances. The studies of cooperative car-following system along these main lines have direct connections to stabilize traffic flow. A usual method is the linear stability analysis providing the conditions for which a small perturbation of steady state will grow over time. For nonlinear effect, the nonlinear equations belong to a particular type of partial differential equations for which the inverse scattering transform allows finding exact solutions [34, 43-44] such as solitary waves, kink and anti-kink waves and triangular waves. Based on OV model and its extended models, many researchers investigated various properties of the traffic flow [45-46].

In general, most current theoretical studies do not consider sufficiently the combined effect of cooperation and different delays. To contribute to the development of traffic flow theory, this paper attempts to derive the analytical conditions reflecting the combined effect of cooperation and delays for the generalized car-following model. To this end, the contributions of this paper are threefold:

(1) The generalized car-following model with consideration of cooperation and time delays is proposed.

(2) Burgers equation and KdV equation are derived and their solitary solutions are obtained.

(3) The combined effect of cooperation and delays is discussed by mean of analytic and simulative methods.

The rest parts of this paper are organized as follows. The general cooperative car-following model with consideration of cooperation and time delays is introduced in Section 2. Section 3 analyzes the linear stability and obtains general stability criterion. Burgers equation and KdV equation and their solutions are formulated in Section 4. As a case, the extended optimal velocity model with time delays is studied and the time evolutions of traffic wave are investigated by using numerical methods in Section 5. We summarize this work in Section 6.

## 2. Model

With the help of wireless communication networks (DSRC, Wifi, 4G, 5G), the vehicle may obtain more information from multiple vehicles ahead or/and behind. Every single vehicle updates its acceleration (or velocity) based on its leading vehicle and following vehicles within a predetermined range, which is called the communication range shown as Fig. 1(b). For traditional car-following model by reason of the lack of communication, every single vehicle updates its acceleration (or velocity) only based on its immediate preceding vehicle shown in Fig. 1(a). The dynamics of individual vehicle can be described by microscopic traffic flow models, e.g., car-following models. Therefore, the acceleration of cooperative traffic dynamic model will be determined as a function of its current velocity and the information of the speed and headways of its preceding vehicles and followers within the communication range. As illustrated in Fig. 1(b), the traffic dynamics of cooperative car-following model should be written as:

$$\ddot{x}_n(t) = f[v_n(t), \Gamma_1\underset{j\in N_1(t)}{(\Delta x_{n,j}(t))}, \Gamma_2\underset{j\in N_2(t)}{(\Delta v_{n,j}(t))}] \tag{1}$$

Where $x_n$ and $v_n$ are location and velocity of vehicle *n*. $\Delta x_{n,j}(t)$ and $\Delta v_{n,j}(t)$ are its space headway, and relative velocity with respect to its neighboring vehicle *j*, respectively. $N_1(t)$ and $N_2(t)$ denote the communication topologies. $\Gamma_1$ and $\Gamma_2$ describe the corresponding control strategies.

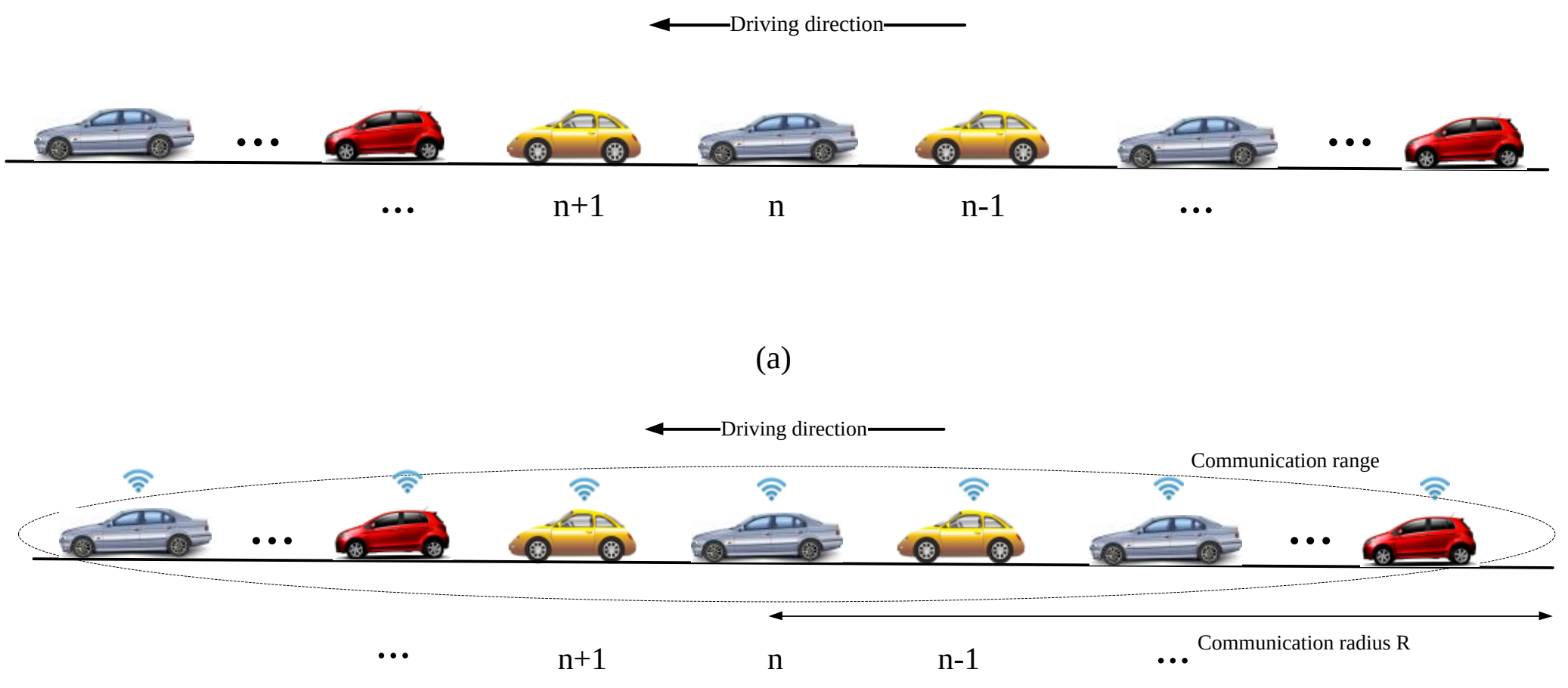

(a)

(b)

**Fig. 1.** Traffic pattern: (a) Traditional car-following system; (b) Cooperative car-following system.

Considering multiple delays, the acceleration of vehicle $n$ can be represented by integrating all the information with the weighting coefficients:

$$\begin{aligned}\ddot{x}_n(t) &= f[v_n(t-\tau_1), \Gamma_1(\underset{j\in N_1(t)}{\Delta x_{n,j}}(t-\tau_2)), \Gamma_2(\underset{j\in N_2(t)}{\Delta v_{n,j}}(t-\tau_3))] \\ &= f[v_n(t-\tau_1), \sum_{j=-k'}^{k} \varphi_j \Delta x_{n+j}(t-\tau_2), \sum_{j=-k'}^{k} \psi_j \Delta v_{n+j}(t-\tau_3)]\end{aligned} \tag{2}$$

Where $k'$ and $k$ is the vehicles number of communication radius. $k'+k$ is the vehicles number within the communication range of vehicle $n$. $\tau_1, \tau_2, \tau_3$ represent reaction delays to different stimuli. The terms with delays on the right-hand side of model (2) derive from reaction times. Due to different levels of perception, delay of host vehicle is different from delay of relative motion sensing to its preceding vehicles and following vehicles. So, we set $\tau_1 \neq \tau_2, \tau_2 = \tau_3$.

$\varphi_j$ and $\psi_j$ are the weighed coefficients of cooperative relation that are supposed to define the importance of the interaction between vehicle $n$ and its surrounding vehicles within its communication range. In realistic traffic flow situations, the downstream influences are more important than the upstream influences for driving behavior. In other words, the influence of driving behavior is asymmetrical. Assume that $\varphi_j^+ \geq \varphi_j^-$ for arbitrary *j*. Cooperation coefficients are chosen as following:

$$\varphi_j = \begin{cases} \beta\varphi_j^+ & \text{for } j \in N^+ \\ (1-\beta)\varphi_j^- & \text{for } j \in N^- \\ 0 & \text{otherwise} \end{cases} \tag{3}$$

where $N^+ = \{n+j \mid j = 1, 2, ..., k\}$ and $N^- = \{n-j \mid j = 1, 2, ..., k'\}$ are the sets of forward and backward considered vehicles, respectively, with $0 \leq \beta \leq 1$ (will be taken as $\beta \geq 0.5$).

In proposed model, $\varphi_j$ and $\psi_j$ satisfy the following two assumptions:

(1) A monotonous decreasing function with $j$. For forward considered vehicles( $j = 1, 2, 3, ..., k$ ), $\varphi_k < ... < \varphi_3 < \varphi_2 < \varphi_1$ and $\psi_k < ... < \psi_3 < \psi_2 < \psi_1$. For backward considered vehicles( $j = -1, -2, -3, ..., -k'$ ), $\varphi_{-k'} < ... < \varphi_{-3} < \varphi_{-2} < \varphi_{-1}$ and $\psi_{-k'} < ... < \psi_{-3} < \psi_{-2} < \psi_{-1}$.

(2) We assume weighed coefficients of cooperative relation are same, that is $\varphi_j = \psi_j$.

(3) $\varphi_j$ is defined as follows [13,42]:

$$\sum_{j=-k'}^{k}\varphi_j = 1,\quad \varphi_j^- = \begin{cases} \dfrac{2}{3^j} & \text{if } j \neq k \\ \dfrac{1}{3^{j-1}} & \text{if } j = k \end{cases},\quad \varphi_j^- = \begin{cases} \dfrac{2}{3^{|j|}} & \text{if } j \neq k' \\ \dfrac{1}{3^{|j-1|}} & \text{if } j = k' \end{cases} \tag{4}$$

The effects of the forward/backward considered vehicles act on the host vehicle will reduce with the increase of distance since the interactive relation decays.

## 3. Linear stability

The traffic flow scenario is a linear spatial configuration of the vehicles (Fig.1). We are interested in determining the evolving processes of traffic flow near steady state by introducing disturbance.

The uniform flow are time independent both the velocities and the headways:

$$\ddot{x}_n = 0,\ \ \Delta x_n = \Delta x_n^* \text{ and } \dot{x}_n = v_n^* \tag{5}$$

For identical vehicles, $\ddot{x}_n = 0,\ \ \Delta x_n = \Delta x^*$ and $\dot{x}_n = v^*$.

When cooperative car-following system reaches the equilibrium state, Eq. (2) satisfies:

$$0 = f(v^*, \Delta x_n^*, 0) \tag{6}$$

For cooperative traffic, we set $\sum_j = \sum_{j=-k'}^{k}$ and let the equilibrium state be characterized by vector:

$$u_n^* = (v_n^*, \Delta x_n^*, 0) \tag{7}$$

with $f(u_n^*) = 0$.

We consider small perturbations to equilibrium state as:

$$x_n(t) = x_n^*(t) + y_n(t) \text{ and } v_n(t) = v_n^*(t) + \dot{y}_n(t) \tag{8}$$

The relation of $u_n(t)$ is introduced as:

$$u_n(t) = u_n^* + \tilde{u}_n(t) \tag{9}$$

where the perturbation $\tilde{u}_n(t)$:

$$\tilde{u}_n(t) = (\dot{y}_n(t), \sum_j \varphi_j \Delta y_{n+j}(t), \sum_j \psi_j \Delta \dot{y}_{n+j}(t)) \tag{10}$$

The formula of $f(\,)$ is deduced by second order Taylor expansion near the equilibrium state as following:

$$f(u_n^* + \tilde{u}_n(t)) = f(u_n^*) + J\tilde{u}_n(t) + \frac{1}{2}(\tilde{u}_n(t))^T H \tilde{u}_n(t) \tag{11}$$

where $J$ and $H$ are the Jacobian matrix and Hessian matrix, respectively

$$J = \frac{\partial f}{\partial \tilde{u}},\ H = \frac{\partial^2 f}{\partial \tilde{u}^2} \tag{12}$$

$$J = [f_{1,n}, f_{2,n}, f_{3,n}],\ H = \begin{bmatrix} h_{11,n} & h_{12,n} & h_{13,n} \\ h_{21,n} & h_{22,n} & h_{23,n} \\ h_{31,n} & h_{32,n} & h_{33,n} \end{bmatrix}$$

where the partial derivatives:

$$f_{1,n} = f_{v,n} = \frac{\partial f(v^*, \Delta x^*, 0)}{\partial v_n},\ f_{2,n} = f_{\Delta x,n} = \frac{\partial f(v^*, \Delta x^*, 0)}{\partial \Delta x_n},\ f_{3,n} = f_{\Delta v,n} = \frac{\partial f(v^*, \Delta x^*, 0)}{\partial \Delta v_n} \tag{13}$$

The partial derivatives of a general cooperative car-following system should satisfy rational driving constraints[14]:

$$f_{1,n} < 0,\ f_{2,n} > 0 \text{ and } f_{3,n} > 0 \tag{14}$$

Since $\Delta \dot{y}_n(t) = v_{n+1}(t) - v_n(t)$ , and hence $\Delta \ddot{y}_n(t) = \dot{v}_{n+1}(t) - \dot{v}_n(t)$ , velocity fluctuations may be eliminated. Eq. (15) is written according to Eq. (2) and Eq.(11), which leads to:

$$\Delta \ddot{y}_n(t) = J(\tilde{u}_{n+1}(t) - \tilde{u}_n(t)) + \frac{1}{2}(\tilde{u}_n(t) + \tilde{u}_{n+1}(t))^T H(\tilde{u}_{n+1}(t) - \tilde{u}_n(t)) \tag{15}$$

where the last term satisfies the symmetry property of the Hessian.

The linearization aims to keep the first-order terms of Eq.(15), we get:

$$\begin{aligned}\Delta \ddot{y}_n(t) = J(\tilde{u}_{n+1}(t) - \tilde{u}_n(t)) = f_{1,n}\Delta \dot{y}_n(t-\tau_1) + f_{2,n}\sum_j \varphi_j(\Delta y_{n+j+1}(t-\tau_2) - \Delta y_{n+j}(t-\tau_2)) \\ + f_{3,n}\sum_j \psi_j(\Delta \dot{y}_{n+j+1}(t-\tau_2) - \Delta \dot{y}_{n+j}(t-\tau_2))\end{aligned} \tag{16}$$

The perturbation is written into the Fourier mode as:

$$\Delta y_n(t) = A_m \exp(i\alpha n + zt) \tag{17}$$

where $\alpha$ is the wave number and $A_m$ is the amplitude of Fourier series.

The following equation of z is obtained from Eq.(16):

$$z^2 - [f_{1,n}e^{-z\tau_1} + f_{3,n}(e^{i\alpha} - 1)\sum_j \psi_j e^{i\alpha j} e^{-z\tau_2}]z - f_{2,n}(e^{i\alpha} - 1)\sum_j \varphi_j e^{i\alpha j} e^{-z\tau_2} = 0 \tag{18}$$

where $z = z_1(ik) + z_2(ik)^2 + z_3(ik)^3 + \dots$, and inserts it into Eq.(18), the first-order and second-order terms of $(i\alpha)$ are collected:

$$\begin{aligned} z_1 &= -\frac{f_{2,n}}{f_{1,n}} \\ z_2 &= \frac{-1}{f_{1,n}}[f_{2,n}\sum_j \varphi_j(j + \frac{1}{2}) - (\frac{f_{2,n}}{f_{1,n}})^2 - \frac{f_{2,n}^2}{f_{1,n}}\tau_1 - \frac{f_{2,n}f_{3,n}}{f_{1,n}}\sum_j \psi_j + \frac{f_{2,n}^2}{f_{1,n}}\sum_j \varphi_j\tau_2]\end{aligned} \tag{19}$$

If $z_2$ is negative, the uniform steady state becomes unstable for long-wavelength modes. If $z_2$ is positive, the uniform steady state becomes stable. Thus, for small disturbances of long wavelength, the uniform traffic flow is stable if

$$\frac{-f_{2,n}}{f_{1,n}^3}[\sum_j \varphi_j(j + \frac{1}{2})f_{1,n}^2 - f_{2,n} - f_{1,n}f_{3,n}\sum_j \psi_j + f_{1,n}f_{2,n}(\sum_j \varphi_j\tau_2 - \tau_1)] > 0 \tag{20}$$

Stability criterion (20) indicates that cooperative mode contributes positively to stabilizing traffic flow while delays contribute negatively to destabilizing traffic flow. According to inequation (20), the delays associated with the relative speed do not contribute to linear stability criterion, which were also proved [15]. However, it could contribute to nonlinear effect that will be investigated in next section.

For cooperative optimal velocity (OV) model with time delays, its mathematical dynamics formulation is presented:

$$\frac{dv_n(t)}{dt} = \kappa[V(\sum_j \varphi_j \Delta x_{n+j}(t-\tau_2)) - v_n(t-\tau_1)] + \lambda\sum_j \varphi_j \Delta v_{n+j}(t-\tau_2) \tag{21}$$

where $\kappa$ and $\lambda$ are sensitivity of driver and gain coefficient, respectively.

$V(\cdot)$ is the optimal velocity function, which depends on the headway $\Delta x_n$. The optimal velocity function is calibrated b with respect to the empirical data [47]:

$$V(\Delta x) = V_1 + V_2 \tanh[C_1(\Delta x - l_c) - C_2] \tag{22}$$

where $V_1 = 6.75m/s, V_2 = 7.91m/s, C_1 = 0.13m^{-1}$, and $C_2 = 1.57$. $l_c$=5$m$ is the length of the vehicles. In addition, we set the parameters $\kappa = 0.8, \lambda = 0.5$ .

For the specifications of the extended OV model above, it is straightforward to show based on Eq.(13):

$$f_{1,n} = -\kappa,\ f_{2,n} = \kappa V',\ f_{3,n} = \lambda \tag{23}$$

$$h_{jk,n} = \begin{cases} \kappa V'', & \text{for } j = 2, k = 2 \\ 0, & \text{otherwise} \end{cases} \tag{24}$$

Where $V' = \frac{\partial V}{\partial \Delta x}|_{\Delta x^*}$ , and $V'' = \frac{\partial^2 V}{\partial \Delta x^2}|_{\Delta x^*}$ .

If $\tau_1 = 0, \tau_2 = 0$ , the stability condition is derived in [13]:

$$\frac{(2V' - \lambda \sum_j \psi_j)}{\sum_j \varphi_j (2j+1)} < \kappa \tag{25}$$

If $\tau_1 = 0, \tau_2 \neq 0$ , the stability condition is derived in [34]:

$$\frac{(2V' - \lambda \sum_j \psi_j)}{\sum_j \varphi_j (2j+1) - 2V' \sum_j \varphi_j \tau_2} < \kappa \tag{26}$$

If $\tau_1 \neq 0, \tau_2 \neq 0, j = 1$, the stability condition is derived in [37-38]:

$$\frac{2(V' - \lambda)}{1 + 2V'(\tau_1 - \tau_2)} < \kappa \tag{27}$$

If $\tau_1 \neq 0, \tau_2 \neq 0$ , the stability condition is obtained according to inequalities (20)and (14):

$$\frac{(2V' - \lambda \sum_j \psi_j)}{\sum_j \varphi_j (2j+1) + 2V'\tau_1 - 2V' \sum_j \varphi_j \tau_2} < \kappa \tag{28}$$

Eq.(28) describes the stability relation of delays and cooperation.

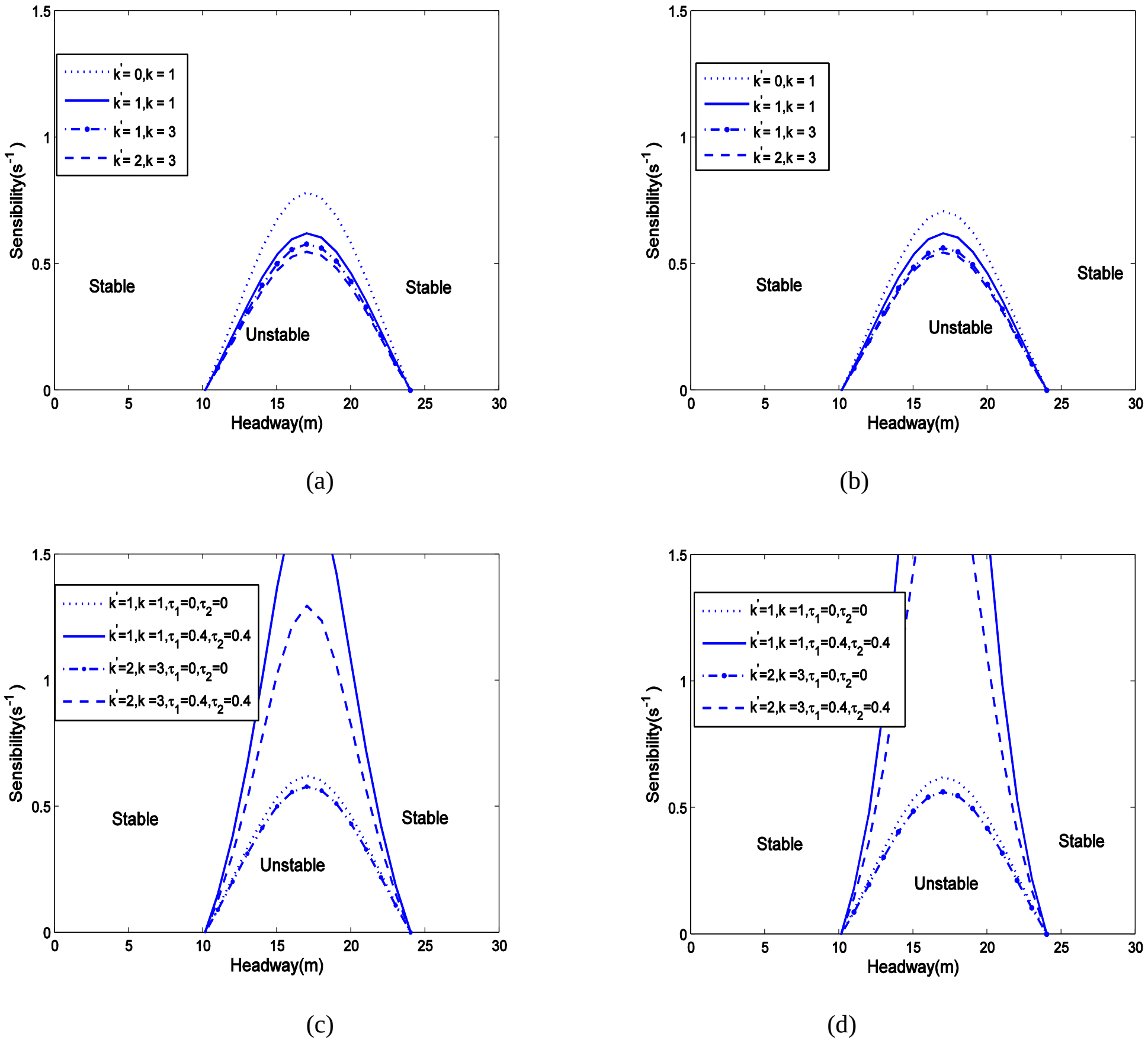

**Fig.2.** Linear stability diagrams of the extended OV model with time delays.

Fig.2 shows that: (a) the weight of asymmetric effect without delays is $\beta = 0.5$, (b) the weight of asymmetric effect without delays is $\beta = 0.7$, (c) the weight of asymmetric effect with delays is $\beta = 0.5$, (d) the weight of asymmetric effect with delays is $\beta = 0.7$. Compared to Fig. 2(a) and (b), cooperative behavior contributes to stabilizing traffic flow, which is consistent with findings in some recent microscopic models and macroscopic models. Meanwhile, the weight of asymmetric effect impacts the stability of traffic flow. Fig.2 (c) and (d) indicate that delays lead to traffic instability, while cooperation effect could stabilize traffic flow. According to analytical and numerical results, delays and cooperation are model-dependent presented in Eq.(28) and Fig.2(c) and (d).

The string instability describes the causes of the oscillations and traffic waves. Therefore, string stability is particularly useful for characterizing disturbance attenuation. Moreover, a chain of vehicles will propagate sufficient growth to trigger nonlinear effects, e.g. solitary wave. The relationship of delays and nonlinear effects is investigates in Section 4.

## 4. Nonlinear analyses

The idea of deriving the nonlinear equation was firstly introduced for the OV model [43-44], and it was developed by mathematicians and physicists in later studies.

We generalize these developments for a class of cooperative car-following models. Burgers equation and KdV equation will be investigated in the weak nonlinear regime and some relations are derived.

Starting from Eq.(15), we set:

$$\bar{u}_{n+1} - \bar{u}_n = \begin{pmatrix} \dot{y}_{n+1}(t-\tau_1) - \dot{y}_n(t-\tau_1) \\ \sum_j \varphi_j(\Delta y_{n+j+1}(t-\tau_2) - \Delta y_{n+j}(t-\tau_2)) \\ \sum_j \psi_j(\Delta \dot{y}_{n+j+1}(t-\tau_2) - \Delta \dot{y}_{n+j}(t-\tau_2)) \end{pmatrix} \tag{29}$$

$$\bar{u}_{n+1} + \bar{u}_n = \begin{pmatrix} \dot{y}_{n+1}(t-\tau_1) + \dot{y}_n(t-\tau_1) \\ \sum_j \varphi_j(\Delta y_{n+j+1}(t-\tau_2) + \Delta y_{n+j}(t-\tau_2)) \\ \sum_j \psi_j(\Delta \dot{y}_{n+j+1}(t-\tau_2) + \Delta \dot{y}_{n+j}(t-\tau_2)) \end{pmatrix} \tag{30}$$

The boundary conditions must be specified. Perturbations are introduced to an equilibrium traffic regime characterized by velocity and headway such that $f(v_n^*, \Delta x_n^*, 0) = 0$. This regime is established maintaining the following boundary conditions. The expression is get as follows:

$$\dot{x}_n = -\Delta\dot{x}_n + \dot{x}_{n+1} = -\Delta\dot{x}_n - \Delta\dot{x}_{n+1} + \dot{x}_{n+2} = \cdots = -\sum_{w=0}^{P_n-1} \Delta\dot{x}_{n+w} + v_n^* \tag{31}$$

where $P_n$ is the total number of vehicles ahead of vehicle *n*. Since $\dot{x}_n = \dot{x}_n^* + \dot{y}_n$, we can rewrite:

$$\dot{y}_n = -\sum_{w=0}^{P_n-1} \Delta\dot{y}_{n+w} \tag{32}$$

Which leads to

$$\dot{y}_n + \dot{y}_{n+1} = -\sum_{w=0}^{P_n-1} \Delta\dot{y}_{n+i} - \sum_{w=0}^{P_{n+1}-1} \Delta\dot{y}_{n+1+w} = -2\sum_{w=0}^{P_n-1} \Delta\dot{y}_{n+w} - \Delta\dot{y}_n \tag{33}$$

The neutral stability is described by $\theta(f_1, f_2, f_3) = -\sum_j \varphi_j (j+\frac{1}{2}) f^2{}_{1,n} + f_{2,n} + f_{1,n} f_{3,n} \sum_j \psi_j - f_{1,n} f_{2,n} (\sum_j \varphi_j \tau_2 - \tau_1)$ . For the sake of simplification, we set $f_{1,n} = f_1, f_{2,n} = f_2, f_{3,n} = f_3$ and $\psi_j = \varphi_j$ . The small scaling parameter $\varepsilon$ is introduced as:

$$\theta(f_1, f_2, f_3) = \varepsilon^2 \tag{33}$$

This ensures that $\theta(f_1, f_2, f_3)$ evolves at the vicinity of the neutral stability.

### 4.1The Burgers equation and its solutions

In this subsection, we consider the wave of traffic flow is long wave patterns from the scale of the coarse graining. By using the reductive perturbation method, Burgers equation is derived in the stable region from Eq.(15). According to multiple scale methods, we introduce multiple scales: space variable *n* and time variable *t*, slow variables *X* and *T* for $0 < \varepsilon \ll 1$.

$$X = \varepsilon(n + bt), T = \varepsilon^2 t, \quad \text{for } 0 < \varepsilon \ll 1 \tag{34}$$

The perturbation is set as:

$$\Delta y_n(t) = \varepsilon R(X, T) \tag{35}$$

where $b$ is an arbitrary constant that will be specified later. According to Eq.(34) and Eq.(35), three derivations are needed: the time derivation, the shift of vehicles from the current position and the Taylor expansion of delays as following:

$$\frac{\partial}{\partial t} = b\varepsilon\partial_X + \varepsilon^2 \partial_T \tag{36}$$

$$R(X_{n+j},T)=\sum_{q=0}^{Q}\frac{(j\varepsilon)^q}{q!}\partial_X^q R(X_n,T) \tag{37}$$

$$R(t-\tau)=R+(-b\tau\varepsilon\partial_X-\tau\varepsilon^2\partial_T)R \tag{38}$$

Substituting Eqs.(29)-(30), and (34)-(38) into Eq.(15) and expanding to the third order of $\varepsilon$ , we obtain the following nonlinear partial differential equation:

$$d_1\partial_X R\varepsilon^2+[d_2\partial_T R+d_3R\partial_X R+d_4\partial_X^2 R]\varepsilon^3=0 \tag{39}$$

where

$$\begin{aligned}
d_1&=bf_1+f_2\\
d_2&=f_1\\
d_3&=h_{12}b+h_{22}\\
d_4&=-b^2+f_2\sum_j\varphi_j(\frac{2j+1}{2})+bf_3-b^2f_1\tau_1-bf_2\tau_2
\end{aligned} \tag{40}$$

Setting $d_1=0$ , that is $b=-f_2/f_1$ , the second-order terms of $\varepsilon$ is eliminated in Eq. (39). We obtain the following partial differential equation:

$$d_2\partial_T R+d_3R\partial_X R+d_4\partial_X^2 R=0 \tag{41}$$

Therefore, Eq.(41) is just the Burgers equation. If $R(X,0)$ is of compact support. Then, the solution $R(X,T)$ of Eq.(41) consists of a series of triangle shock waves analyzed in the following case study.

For extended optimal velocity model, we obtain the Burgers equation by substituting into Eqs.(40) and (41):

$$\begin{aligned}
d_1&=-b\kappa+\kappa V'\\
d_2&=-\kappa\\
d_3&=\kappa V''\\
d_4&=-b^2+\kappa V'\sum_j\varphi_j(\frac{2j+1}{2})+b\lambda+b^2\kappa\tau_1-b\kappa V'\tau_2
\end{aligned} \tag{42}$$

When $V''<0$ , the coefficient $d_3$ is negative in the stable region and Eq.(20) is satisfied. The solution of Burgers equation is formulated as:

$$R(X,T)=\frac{1}{|V''|T}[X-\frac{\eta_{n+1}+\eta_n}{2}]-\frac{\eta_{n+1}-\eta_n}{2|V''|T}\tanh\left[B\frac{(\eta_{n+1}-\eta_n)(X-\xi_n)}{4|V''|T}\right] \tag{43}$$

where $B=[\sum_j\varphi_j(\frac{2j+1}{2})-\frac{V'}{\kappa}+\frac{\lambda}{\kappa}+(\tau_1-\tau_2)V']V'$ , $\xi_n$ are the coordinates of the wave fronts and $\eta_n$ are the coordinates of the intersections of the slopes with the x-axis( $n=1,2,...,N$ ). From Eq.(43), $R(X,T)$ tends to 0 when $T\to\infty$ , which means any triangle wave expressed by Eq.(41) in stable traffic flow region will evolve to a uniform flow when time is sufficient large. Eq.(43) presents that the triangle wave propagates backward with the propagation velocity $v_P=V'$ , but it propagates forward in the absolute system if it is in the stable region. The propagation speed decreases with the increase of average headway. This phenomenon is shown in Figs. (4) and (7).

## 4.2 The KdV equation and its solutions

In this subsection, we derive the KdV equation to describe the soliton wave near the neutral stability. Nonlinear analysis is conducted to study the slowly varying behavior near the critical point. For extracting slow scales with the space variable *n* and the time variable *t*, the slow variable *X* and *T* are defined as follows:

$$X = \varepsilon(n+bt), T = \varepsilon^3 t, \quad \text{for } 0 < \varepsilon \ll 1 \tag{44}$$

where $b$ is undetermined parameter that will be specified later. The perturbation is set as:

$$\Delta y_n(t) = \varepsilon^2 R(X,T) \tag{45}$$

According to Eqs.(44) and (45), three basic operations are needed: the time derivation, the shift of vehicles from the current position and the Taylor expansion of delays. We obtain Eqs.(46)-(48),

$$\frac{\partial}{\partial t} = b\varepsilon\partial_X + \varepsilon^3\partial_T \tag{46}$$

$$R(X_{n+j},T) = \sum_{q=0}^{Q}\frac{(j\varepsilon)^q}{q!}\partial_X^q R(X_n,T) \tag{47}$$

$$R(t-\tau) = R + (-b\tau\varepsilon\partial_X - \tau\varepsilon^3\partial_T)R \tag{48}$$

Substituting Eqs.(29)-(30) and (44)-(48) into Eq.(15) is expanded to the sixth order of $\varepsilon$. Then, we obtain the following nonlinear partial differential equation as following:

$$\begin{aligned} &g_3\partial_X R\varepsilon^3 + g_4\partial_X^2 R\varepsilon^4 + (g_5\partial_T R + g_6 R\partial_X R + g_7\partial_X^3 R)\varepsilon^5 \\ &+(g_8\partial_T\partial_X R + g_9(\partial_X R)^2 + g_{10}R\partial_X^2 R + g_{11}\partial_X^4 R)\varepsilon^6 = 0 \end{aligned} \tag{49}$$

The expressions can be skillfully obtained by simple identification.

Term $g_3$ :

$$g_3 = (bf_1 + f_2) \tag{50}$$

In order to remove this term, we set $b = -f_2 / f_1$.

Term $g_4$ :

$$g_4 = \frac{f_2}{f_1^2}[-\sum_j \varphi_j (j+\frac{1}{2})f^2_{1,n} + f_{2,n} + f_{1,n}f_{3,n}\sum_j \varphi_j - f_{1,n}f_{2,n}(\sum_j \varphi_j\tau_2 - \tau_1)] \tag{51}$$

The relation of linear stability(20) defines the close proximity to the neutral stability, so the $g_4$ term within the bracket is equal to $\varepsilon^2$. Then, term $g_4$ is of order $\varepsilon^6$ and can be discarded.

The terms of fifth order and the sixth order are formulated as following:

$$
\begin{aligned}
g_5 &= f_1 \\
g_6 &= -h_{21}\frac{f_2}{f_1} + h_{22} \\
g_7 &= f_2\sum_j \varphi_j \frac{3j^2+3j+1}{6} - \frac{f_2 f_3}{f_1}(\frac{1}{2}+\sum_j \varphi_j j) + \tau_2 \frac{f_2^2}{f_1}(\frac{1}{2}+\sum_j \varphi_j j) - \tau_2 \frac{f_2^2 f_3}{f_1^2} \\
g_8 &= 2\frac{f_2}{f_1} + f_3 - 2f_2\tau_1 - f_2\tau_2 \\
g_9 &= \frac{f_2}{2f_1}(-\frac{f_2}{f_1}h_{11} + h_{12})(2S_0+1) - h_{21}\frac{f_2}{f_1}\sum_j \varphi_j (2j+1) \\
&+\frac{1}{2}h_{22}\sum_j \varphi_j (2j+1) + \frac{f_2^2}{f_1^2}h_{31} - h_{32}\frac{f_2}{f_1} - \frac{f_2^2}{f_1^2}h_{21}\tau_2 + \frac{f_2}{f_1}h_{22}\tau_2 \\
g_{10} &= \frac{1}{2}h_{22}\sum_j \varphi_j (2j+1) - h_{23}\frac{f_2}{f_1} - h_{21}\frac{f_2^2}{f_1^2}\tau_1 + \frac{f_2}{f_1}h_{22}\tau_2 \\
g_{11} &= f_2\sum_j \varphi_j \frac{(j+1)^4 - j^4}{4!} - \frac{f_2 f_3}{f_1}\sum_j \varphi_j \frac{3j^2+3j+1}{6} \\
&+\frac{f_2^2}{f_1}\tau_2\sum_j \varphi_j \frac{3j^2+3j+1}{6} - \frac{f_2^2 f_3}{f_1^2}\tau_2(\frac{1}{2}+\sum_j \varphi_j j)
\end{aligned}
\qquad (52)
$$

where $S_Q = \sum_{q=0}^{Q}\sum_{l=1}^{P}(\varepsilon l)^q \frac{1}{q!}\partial_X^q$ .

So, Eq.(49) could be simplified as:

$$
\begin{aligned}
&(g_5\partial_T R + g_6 R\partial_X R + g_7\partial_X^3 R)\varepsilon^5 \\
&+[g_8\partial_T\partial_X R + g_9(\partial_X R)^2 + g_{10}R\partial_X^2 R + g_{11}\partial_X^4 R + \frac{f_2}{f_1^2}\partial_X^2 R]\varepsilon^6 = 0
\end{aligned}
\qquad (53)
$$

Firstly, we consider the terms of fifth order and ignore $O(\varepsilon)$ term. Its cancelation leads to the KdV-type equation as:

$$
g_7\partial_X^3 R + g_6 R\partial_X R + g_5\partial_T R = 0 \qquad (54)
$$

In order to derive the standard KdV equation with higher-order correction, we make the following transformation in Eq.(54):

$$
R = (g_7^{1/3}/g_6)R',\, X = g_7^{1/3}X',\ \text{and}\ T = g_5 T' \qquad (55)
$$

Hence, we have KdV equation and solution $R_0'(X',T')$ is known as:

$$
\partial_{X'}^3 R' + R'\partial_{X'}R' + \partial_{T'}R' = 0 \qquad (56)
$$

$$
R_0'(X',T') = A\mathrm{sech}^2[\sqrt{\frac{A}{12}}(X' - \frac{A}{3}T')] \qquad (57)
$$

Then, the $O(\varepsilon)$ correction is considered. We assume that

$$
R'(X',T') = R_0'(X',T') + \varepsilon R_1'(X',T') \qquad (58)
$$

The perturbation term of Eq.(45) gives the condition of selecting a unique member from the continuous family of KdV solitons:

$$
R_0(X,T) = A\frac{g_7^{1/3}}{g_6}\mathrm{sech}^2[\sqrt{\frac{A}{12}\frac{1}{g_7^{2/3}}}(X - \frac{A}{3}\frac{g_7^{1/3}}{g_5}T)] \qquad (59)
$$

where $A$ is a free parameter. It is the amplitude of soliton solutions of the KdV equation. In order to obtain the value of $A$, the solvability condition is

$$(R'_0, G[R'_0]) = \int_{-\infty}^{+\infty} dX R'_0 G[R'_0] = 0 \tag{60}$$

Eq.(60) must be satisfied, here $G[R'_0]$ is the $O(\varepsilon)$ term in Eq.(49).

The $R'_1(X', T')$ term should be rewritten as:

$$R'_1(X', T') = \frac{1}{g_6}[m_1 \partial_{X'}^4 R' + m_2 (\partial_{X'} R')^2 + +m_3 \partial_{X'}^2 R'^2 + m_4 \partial_{X'}^2 R'] \tag{61}$$

where

$$\begin{aligned} m_1 &= \frac{1}{g_7}(g_{11} - \frac{g_7 g_8}{g_5}) \\ m_2 &= \frac{1}{g_6}(g_9 - \frac{g_6 g_8}{g_5}) \\ m_3 &= \frac{1}{2g_6}(g_{10} - \frac{g_6 g_8}{g_5}) \\ m_4 &= \frac{1}{g_7^{1/3}} \frac{f_2}{f_1^2} \end{aligned} \tag{62}$$

Performing the integration gives the amplitude, the unique equation of the soliton which is derived at the vicinity of the neutral stability. We obtain the value of amplitude $A$[14]:

$$A = m_4 \frac{16}{15}[m_1 \frac{16}{63} + m_2 \frac{64}{105} - m_3 \frac{128}{105}]^{-1} \tag{63}$$

Using Eq.(45), we can now give the expression of the perturbation:

$$\Delta y_n(t) = \mathrm{sgn}(g_6) \Lambda \mathrm{sech}^2 [\sqrt{\frac{|g_6| \Lambda}{12 g_7}} (n - \alpha_s t)] \tag{64}$$

where

$$\Lambda = \theta A \frac{g_7^{1/3}}{|g_6|}, \alpha_s = \frac{3 f_2 + \Lambda |g_6|}{3 f_1} \tag{65}$$

We analyze the evolution of traffic flow with small perturbations under different delays and find the soliton wave near the neutral stability line.

For cooperative OV model, we get the terms according to formulas(23) and (24):

$$
\begin{aligned}
g_5 &= -\kappa \\
g_6 &= \kappa V'' \\
g_7 &= \kappa V' \sum_j \varphi_j \frac{3j^2+3j+1}{6} + V'\lambda(\frac{1}{2}+\sum_j \varphi_j j) - \tau_2 (V')^2 [\kappa(\frac{1}{2}+\sum_j \varphi_j j)+\lambda] \\
g_8 &= -2V' + \lambda - 2\kappa V'\tau_1 - \kappa V'\tau_2 \\
g_9 &= \frac{1}{2}\kappa V'' \sum_j \varphi_j (2j+1) - V'\kappa V''\tau_2 \\
g_{10} &= \frac{1}{2}\kappa V'' \sum_j \varphi_j (2j+1) - V'\kappa V''\tau_2 \\
g_{11} &= \kappa V' \sum_j \varphi_j \frac{(j+1)^4 - j^4}{4!} + V'\lambda \sum_j \varphi_j \frac{3j^2+3j+1}{6} \\
&-(V')^2 \tau_2 [\kappa \sum_j \varphi_j \frac{3j^2+3j+1}{6} + (\frac{1}{2}+\sum_j \varphi_j j)]
\end{aligned} \tag{66}
$$

We obtain the KdV equation with higher-order correction term:

$$
\begin{aligned}
&\partial_{T'} R' + \partial^3_{X'} R' + R'\partial_{X'} R' \\
&+\frac{1}{g_6}[m_1 \partial^4_{X'} R' + m_2 (\partial_{X'} R')^2 + +m_3 \partial^2_{X'} R'^2 + m_4 \partial^2_{X'} R']\varepsilon = 0
\end{aligned} \tag{67}
$$

where

$$
\begin{aligned}
m_1 &= \frac{\kappa[\sum_j \varphi_j \frac{(j+1)^4 - j^4}{4!}] + \lambda \sum_j \varphi_j \frac{3j^2+3j+1}{6} + [\kappa \sum_j \varphi_j \frac{3j^2+3j+1}{6} - \lambda(\frac{1}{2}+\sum_j \varphi_j j)]V'\tau_2}{\kappa \sum_j \varphi_j \frac{3j^2+3j+1}{6} + \lambda(\frac{1}{2}+\sum_j \varphi_j j) - [\kappa(\frac{1}{2}+\sum_j \varphi_j j)+\lambda]V'\tau_2} \\
&- \frac{2V' - \lambda + \kappa V'(2\tau_1 + \tau_2)}{\kappa} \\
m_2 &= (\frac{1}{2}+\sum_j \varphi_j j) - V'\tau_2 - \frac{2V' - \lambda + \kappa V'(2\tau_1+\tau_2)}{\kappa} \\
m_3 &= \frac{1}{2} m_2 \\
m_4 &= \frac{V'}{\kappa}\left\{\kappa V' \sum_j \varphi_j \frac{3j^2+3j+1}{6} + V'\lambda(\frac{1}{2}+\sum_j \varphi_j j) - [\kappa(\frac{1}{2}+\sum_j \varphi_j j)+\lambda]V'^2\tau_2\right\}^{(-\frac{1}{3})}
\end{aligned} \tag{68}
$$

We obtain the value of amplitude *A:*

$$
A = m_4 \frac{16}{15}[m_1 \frac{16}{63} + m_2 \frac{64}{105} - m_3 \frac{128}{105}]^{-1} \tag{69}
$$

Using Eq. (41), now we can give the expression of the perturbation:

$$
\Delta y_n(t) = \Lambda \mathrm{sech}^2[\sqrt{\frac{|g_6|\Lambda}{12 g_7}}(n - \alpha_s t)] \tag{70}
$$

where

$$
\Lambda = \varepsilon^2 A \frac{g_7^{1/3}}{|g_6|}, \alpha_s = \frac{3 f_2 + \Lambda |g_6|}{3 f_1} \tag{71}
$$

The solutions of Burgers equation and KdV equation are obtained, and the jamming transition of traffic flow can be described by these equations.

## 5. Simulations

To validate the results of theoretical analysis, we assume that all vehicles presented by the extended OV model (21) move under the periodic boundary condition. In this scenario, there are $N=100$ vehicles running on a ring road with the length $L=1500$ m and all vehicles are initially traveling at the nominal speed of $V(L/N)$ with the constant headway of $L/N$ presented by Fig. 3. Eq.(22) is adopted as optimal velocity function. Considered vehicles are $k=k'=3$ and the weight of asymmetry $\beta=0.7$. Other parameters are same with the extended OV model above.

Then, we assume the initial disturbance as following:

$$\begin{aligned} & x_1(0)=1\,m;\quad x_n(0)=(n-1)\frac{L}{N} \quad \text{for } n\neq 1 \\ & v_n(0)=V(\frac{L}{N}) \end{aligned} \tag{72}$$

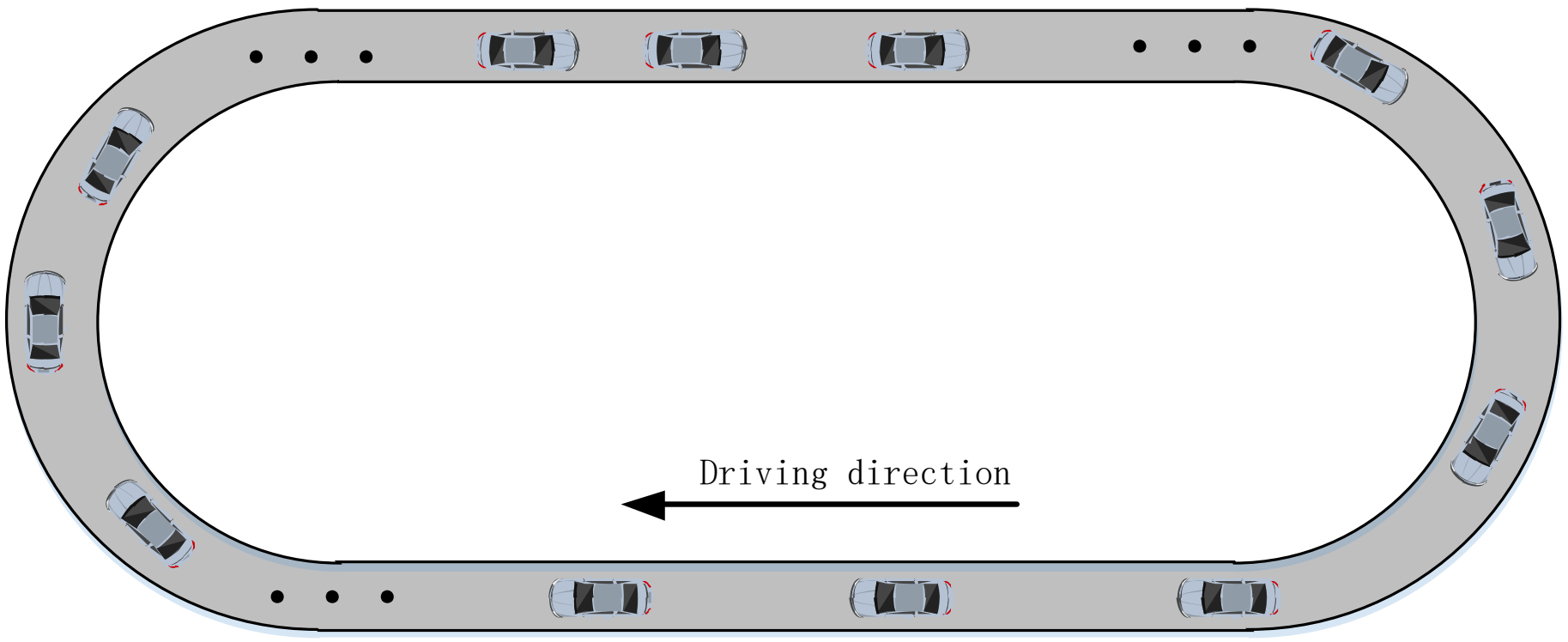

**Fig. 3.** The diagram of ring road

Substituting the values of the parameters into criterion(28), we learn that the initial disturbance is stable. We investigate the influence of different delays from the vehicle dynamics to traffic flow and density waves.

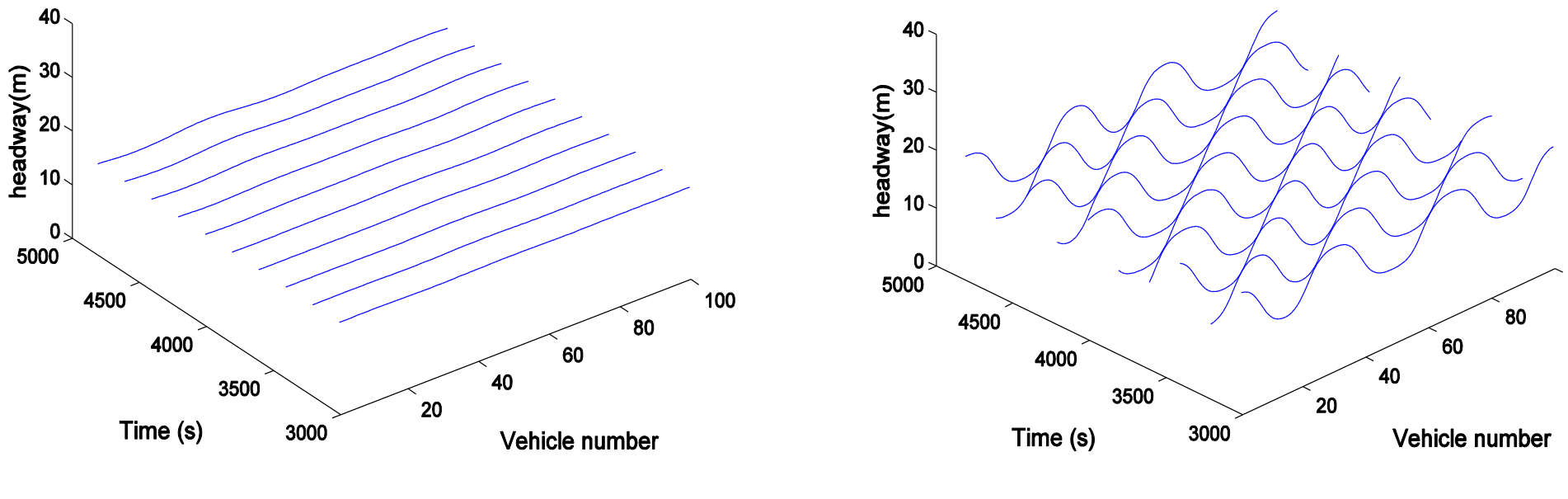

(a) (b)

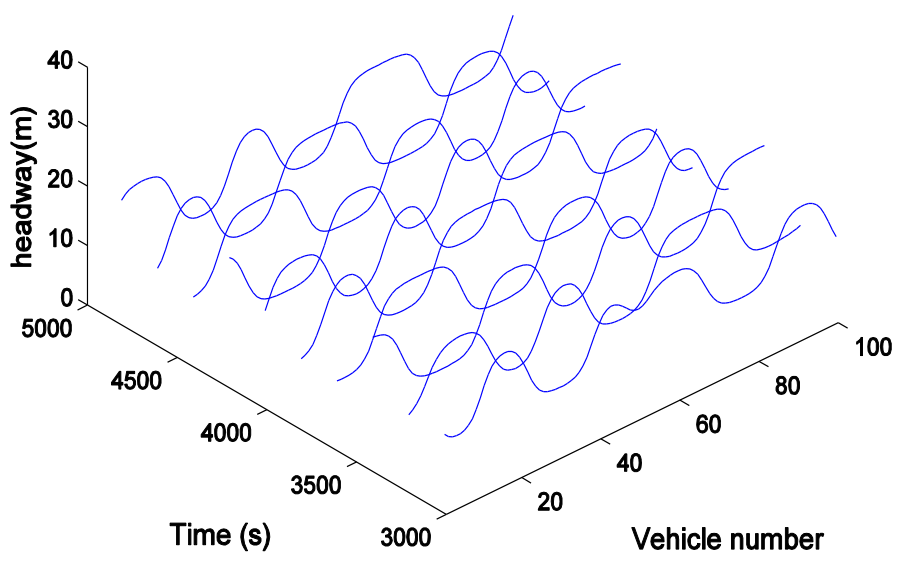

(c)

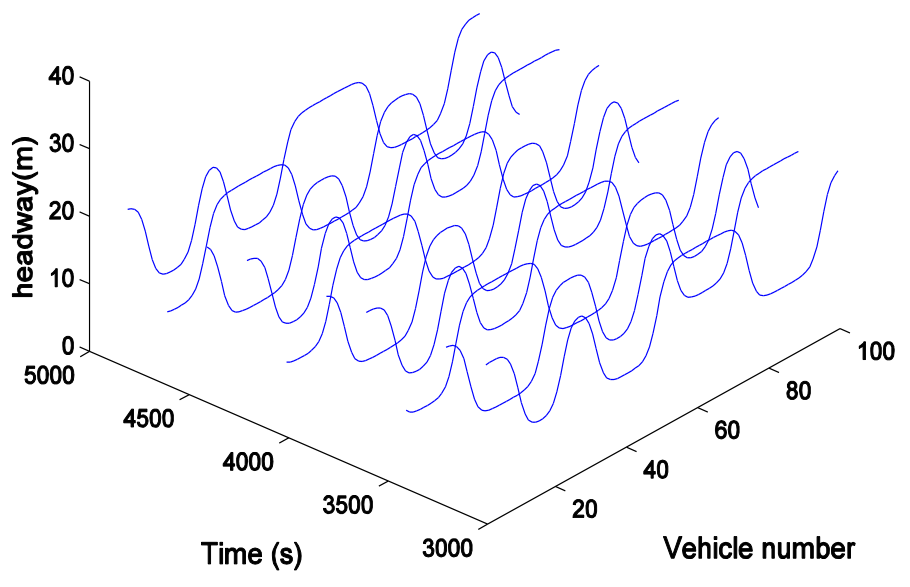

(d)

**Fig.4.** Spatial-temporal evolution patterns of headway for $\tau_1 = 0.2$ : (a) $\tau_2 = 0.1$ , (b) $\tau_2 = 0.2$ , (c) $\tau_2 = 0.3$ , (d) $\tau_2 = 0.4$

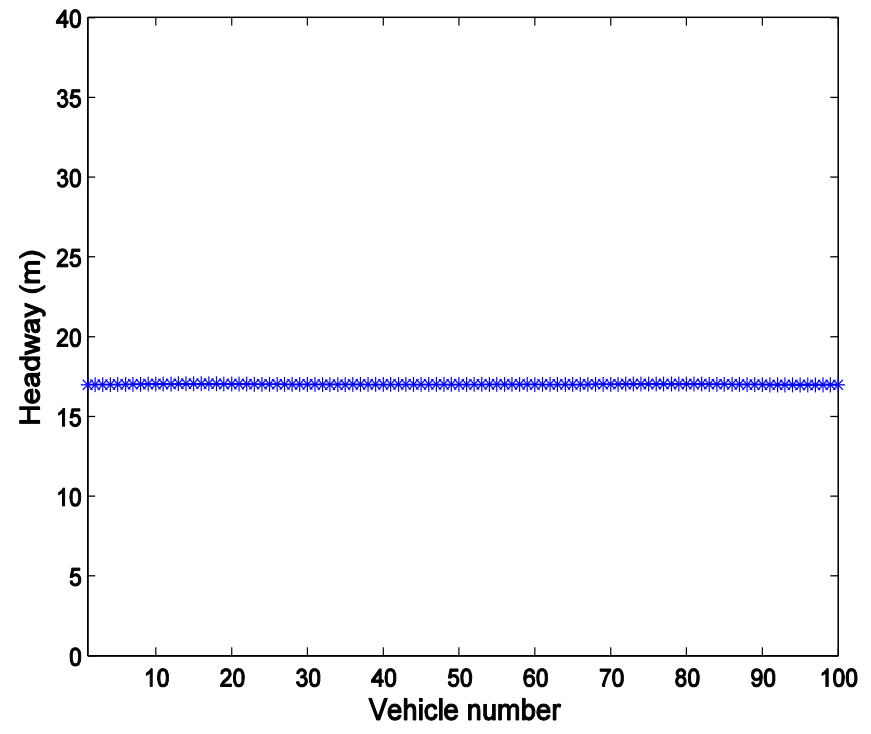

(a)

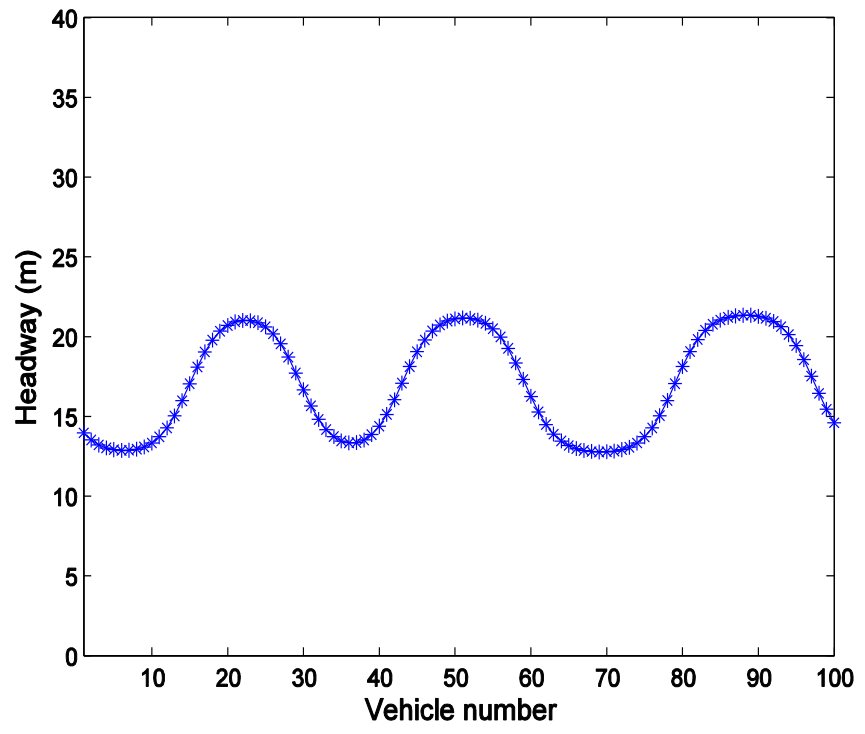

(b)

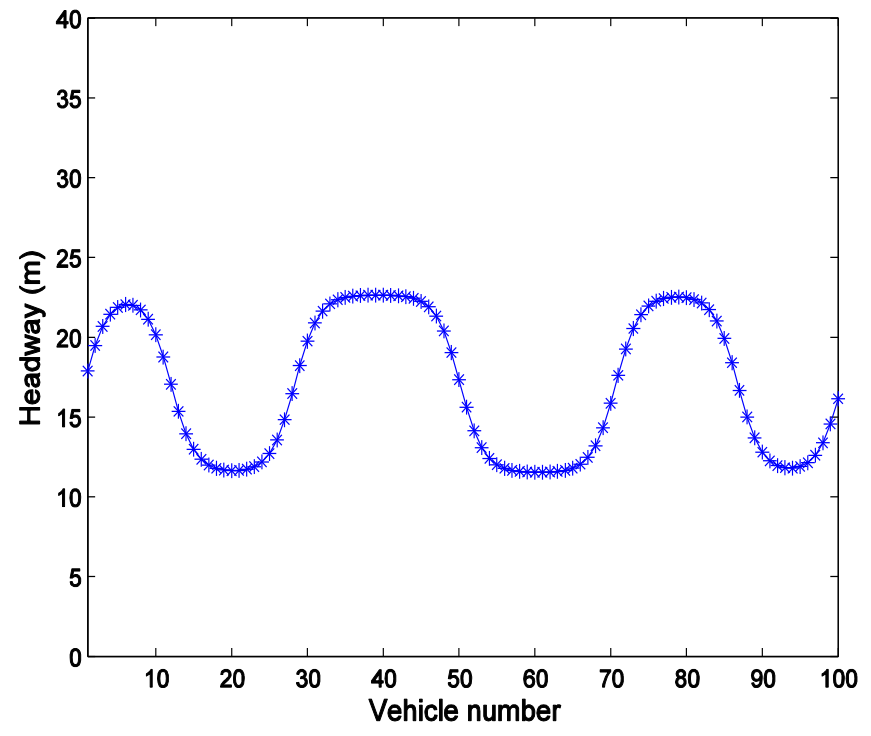

(c)

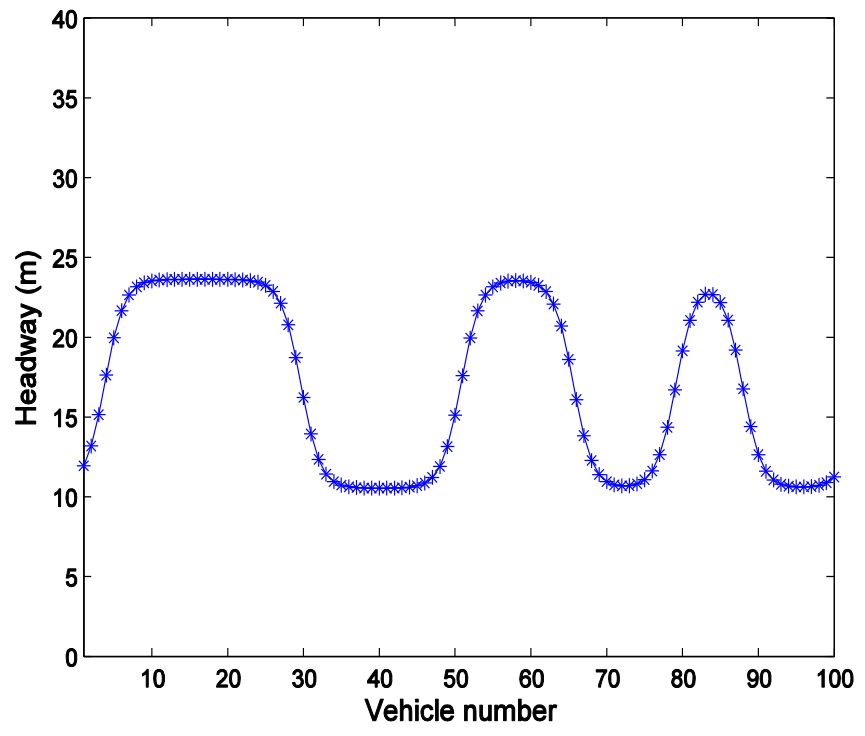

(d)

**Fig.5.** Headway profiles of density wave at *t*=4000*s* correspond to Fig. 4.

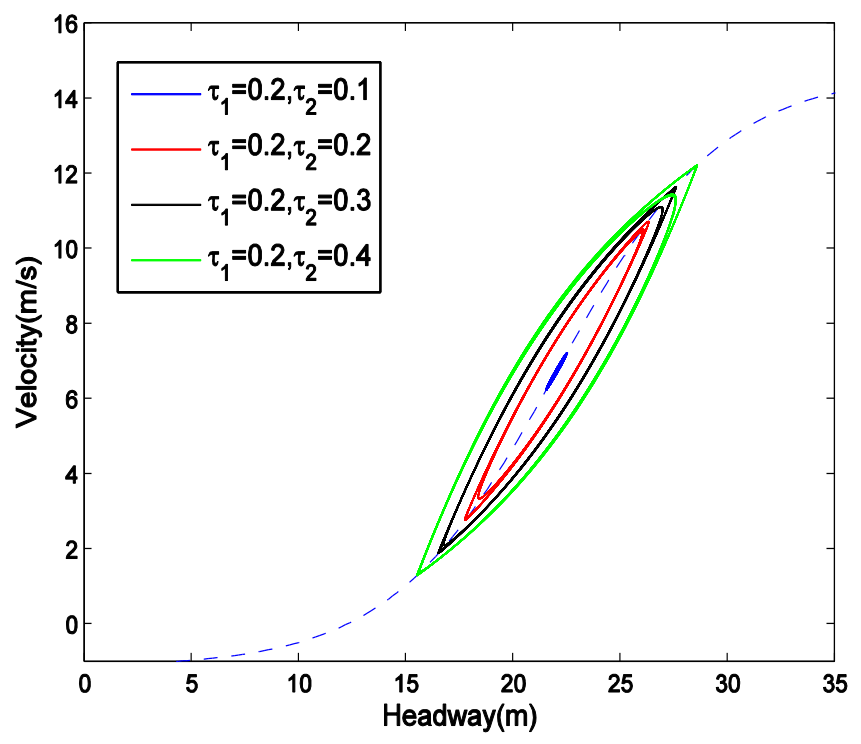

**Fig. 6.** Hysteresis loops considering different delays correspond to Fig. 4.

Fig. 4 shows that spatial-temporal evolution patterns of headway after a sufficiently long time with different $\tau_2$ and constant $\tau_1$. Fig. 4 and Fig. 5 exhibit the time evolution of traffic wave and corresponding headway profile obtained by *t*=4000*s*, respectively. By using linear stability condition (28), traffic flow is stable for $\tau_1 = 0.2$ and $\tau_2 = 0.1$. So, small disturbances will be attenuated, traffic flow is uniform over the whole space shown in Fig. 4(a), Fig. 5(a). When $\tau_1 = 0.2$ is constant, for $\tau_2 = 0.2$, $\tau_2 = 0.3$ and $\tau_2 = 0.4$, those don't satisfy linear stability condition, corresponding to Fig. 4(b),(c),(d) and Fig.5(b),(c),(d).

Meanwhile, hysteresis loops will expand with increasing $\tau_2$ shown in Fig.6, which corresponds to traffic flow instability. If the traffic flow is stable, hysteresis loops will not be generated in phase space, and there will be only a point on the optimal velocity curve instead. This indicates that traffic jams will be worse when increasing $\tau_2$.

So small disturbances will be amplified; then, those perturbations will propagate sufficient growth along platoon to trigger nonlinear effects; finally, traffic flow will occur oscillations and traffic wave. From Fig. 4(b) to (d), the triangular shock waves, solitary waves and oscillations (stop-and-go behavior) will arise in order.

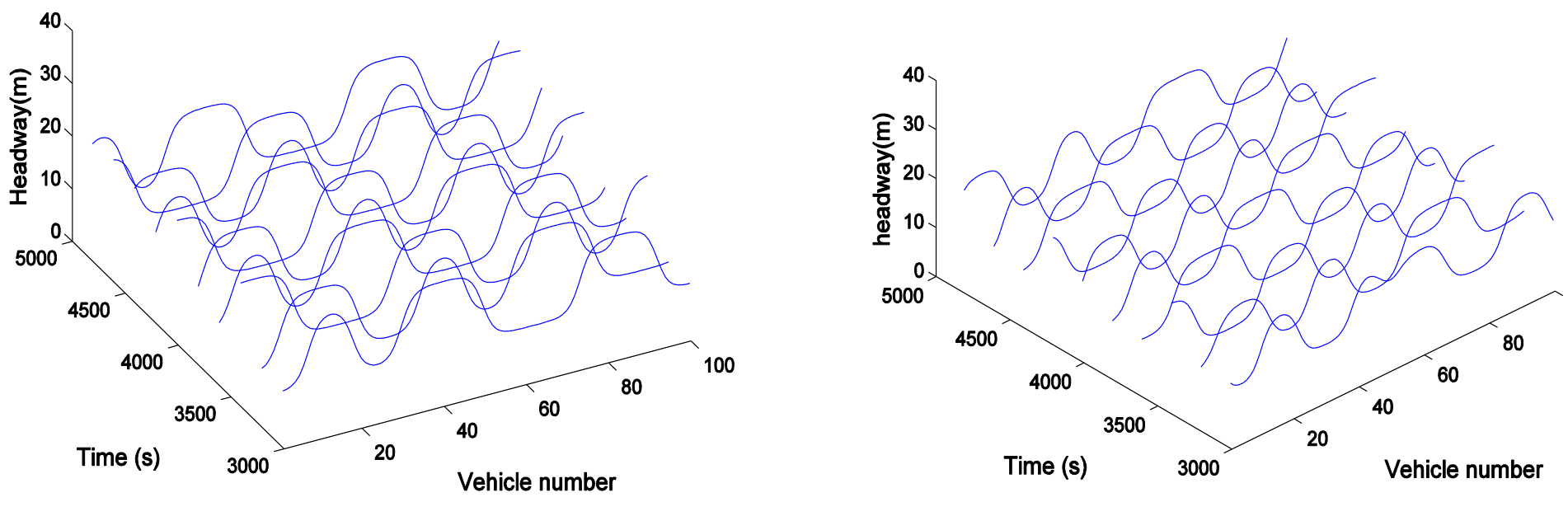

(a) (b)

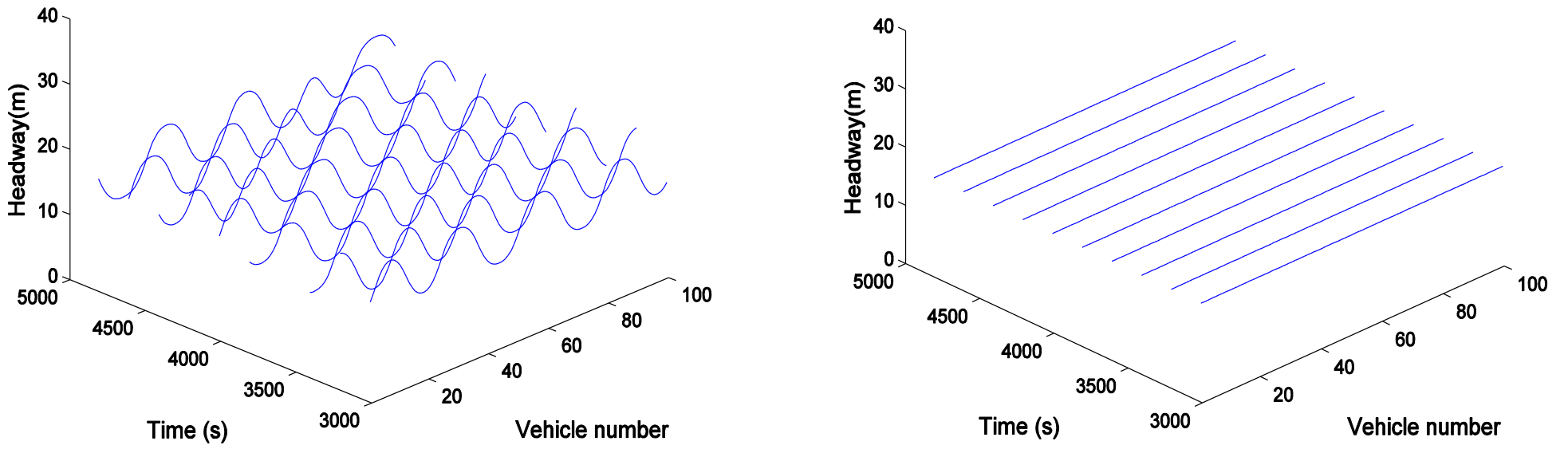

(c) (d)

**Fig.7.** Spatial-temporal evolution patterns of headway for $\tau_2 = 0.3$ : (a) $\tau_1 = 0.1$, (b) $\tau_1 = 0.2$ , (c) $\tau_1 = 0.3$ , (d) $\tau_1 = 0.4$

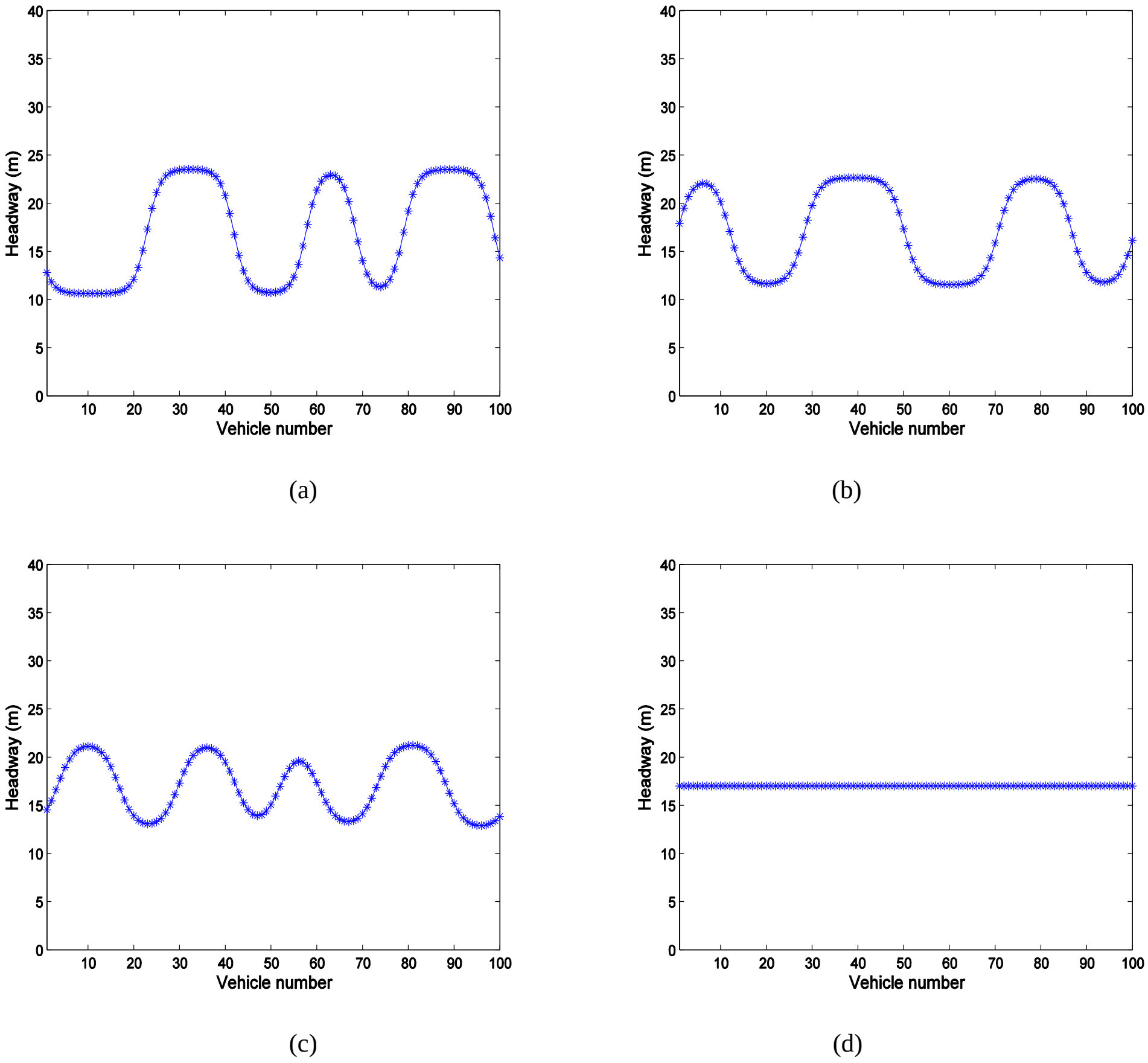

**Fig.8.** Headway profiles of density wave at $t$=4000$s$ correspond to Fig. 7.

Fig. 7 shows that spatial-temporal evolution patterns of headway after a sufficiently long time with different $\tau_2$ and constant $\tau_1$ . Fig. 7 and Fig. 8 exhibit the time evolution of traffic wave and corresponding headway profile obtained by $t$=4000$s$, respectively. For $\tau_1 = 0.1$ , $\tau_1 = 0.2$ and $\tau_1 = 0.3$ , small disturbances will be amplified; then, small disturbances will propagate sufficient growth to trigger nonlinear effects; finally, traffic flow will occur oscillations and traffic wave. The triangular shock waves, solitary waves and oscillations will arise in order corresponding to Fig. 7(a),(b),(c) and Fig.8(a),(b),(c). By using linear stability condition (28), traffic flow is stable for $\tau_1 = 0.4$ and $\tau_2 = 0.3$ .

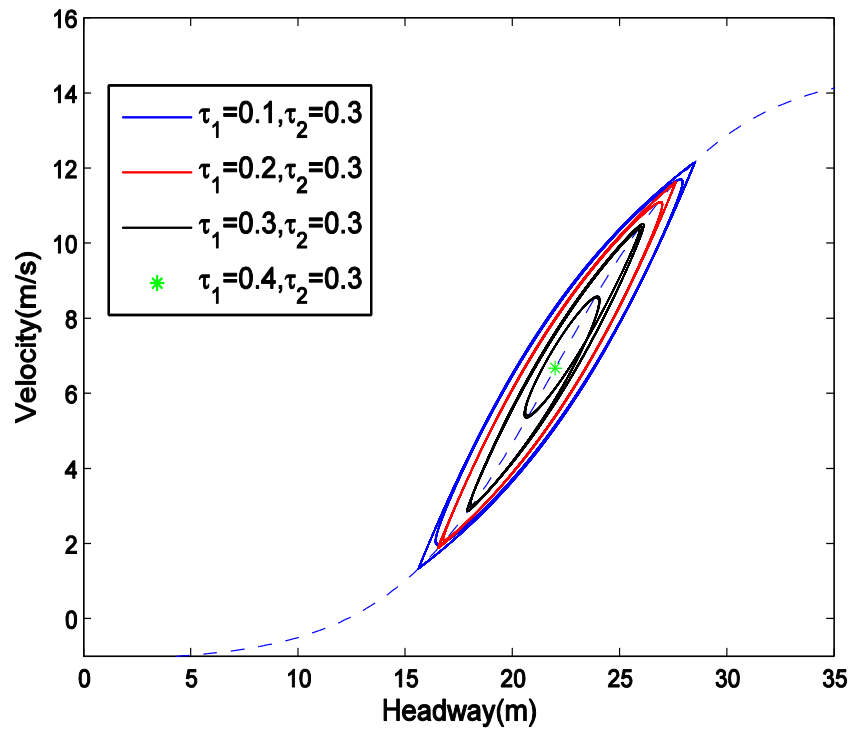

**Fig. 9.** Hysteresis loops considering different delays correspond to Fig. 7.

Hysteresis loops will expand with increasing $\tau_1$ shown in Fig.6. When delay of $\tau_1$ increases, the traffic flow will be stable and hysteresis loops will not be generated in phase space. This indicates that traffic jams will be stabilized when delay of $\tau_1$ increases.

So, small disturbances will be attenuated, traffic flow is uniform over the whole space shown in Fig. 7(d), Fig. 8(d). This indicates that increasing appropriately driver's reaction time sensing to host velocity is helpful to reduce the nonlinear effects a certain extent.

## 6. Conclusions

In this paper, we present the general car-following model with consideration of cooperation and time delays in cyber-physical perspective. General linear stability condition is investigated and stability criterions are formulated. It is found that delays and cooperation are model-dependent. Then the relation of cooperation and time delays is discussed. The comprehensive methods are generalized to derive Burgers equation and KdV equation for generic car-following model. Meanwhile, their solitary wave solutions and constraint conditions are obtained. The property of the cooperative OV model which estimates the impact of delays about the evolution of traffic congestion is investigated by using both analytic and numerical methods. The oscillations and stop-and-go waves are studied. The numerical results show that traffic jams are suppressed when the cooperation is considered. Moreover, delays of sensing to relative motion are easy to trigger the traffic waves; delays of sensing to host vehicle are beneficial to relieve the instability effect a certain extent.

## Acknowledgments

This work was supported by the National Natural Science Foundation of China (Grant No. 61573075, Grant No. 71201178), the National Key R&D Program (Grant No.2016YFB0100904), the Natural Science Foundation of Chongqing (Grant No. cstc2017jcyjBX0001), the Fundamental Research Funds for the Central Universities (Grant No. 106112016CDJXY170002) and the Foundation for High-level Talents of Chongqing University of Art and Sciences(Grant No. 2017RJD13).